\documentclass[aps,twocolumn,english,superscriptaddress,citeautoscript,showkeys,preprintnumbers,amsmath,amssymb,floatfix,footinbib,prb]{revtex4-2}
\usepackage{graphicx}
\usepackage{xcolor}
\usepackage{soul}
\usepackage{amssymb,amsmath}
\usepackage{hyperref}
\hypersetup{colorlinks=true,linkcolor=red,citecolor=blue}
\usepackage[utf8]{inputenc}
\usepackage[english]{babel}
\usepackage{txfonts}
\DeclareUnicodeCharacter{0303}{HERE!HERE!}

\newcommand{\Tc}{$T_\text{c}$}

\begin{document}

\title{Nevanlinna Analytic Continuation for Migdal-Eliashberg Theory}

\author{D. M. Khodachenko}
\affiliation{Institute of Theoretical and Computational Physics, Graz University of Technology, NAWI Graz, 8010, Graz, Austria}
\author{R. Lucrezi}
\affiliation{Institute of Theoretical and Computational Physics, Graz University of Technology, NAWI Graz, 8010, Graz, Austria}
\author{P. N. Ferreira}
\affiliation{Institute of Theoretical and Computational Physics, Graz University of Technology, NAWI Graz, 8010, Graz, Austria}
\affiliation{Computational Materials Science Group (ComputEEL/MatSci), Universidade de S\~ao Paulo, Escola de Engenharia de Lorena, DEMAR, Lorena, Brazil}
\author{M. Aichhorn}
\affiliation{Institute of Theoretical and Computational Physics, Graz University of Technology, NAWI Graz, 8010, Graz, Austria}
\author{C. Heil}
\email[Corresponding author: ]{christoph.heil@tugraz.at}
\affiliation{Institute of Theoretical and Computational Physics, Graz University of Technology, NAWI Graz, 8010, Graz, Austria}

\begin{abstract}
In this work, we present a method to reconstruct real-frequency properties from analytically continued causal Green's functions within the framework of Migdal-Eliashberg (ME) theory for superconductivity. ME theory involves solving a set of coupled equations self-consistently in imaginary frequency space, but to obtain experimentally measurable properties like the spectral function and quasiparticle density of states, it is necessary to perform an analytic continuation to real frequency space. Traditionally, the ME Green's function is decomposed into three fundamental complex functions, which are analytically continued independently. However, these functions do not possess the causal properties of Green's functions, complicating or even preventing the application of standard methods such as Maximum Entropy. Our approach overcomes these challenges, enabling the use of various analytic continuation techniques that were previously impractical. We demonstrate the effectiveness of this method by combining it with Nevanlinna analytic continuation to achieve accurate real-frequency results for ME theory, which are directly comparable to experimental data, with applications highlighted for the superconductors MgB$_2$ and LaBeH$_8$.
\end{abstract}

\date{\today}

\pacs{}

\maketitle

\section{Introduction}
Temperature-dependent quantum field theory via Green's functions is a state-of-the-art approach to theoretical condensed matter physics. Solving many-body Green's function equations directly on the real frequency axis is possible, but numerically very challenging, due to the existence of poles on the real axis, requiring the evaluation of principal value integrals. 
However, these equations simplify significantly when formulated in imaginary, i.e. Matsubara frequency space. 
By doing so, the problem becomes markedly simpler both analytically and numerically, as it leads to well-defined integrals and discrete Matsubara frequency sums~\cite{Nolting}, reducing the numerical cost of calculations considerably compared to a direct evaluation in otherwise continuous real frequency space. 

The disadvantage that comes with this is that in order to be able to compare to real frequency properties obtained in experiment, such as angle-resolved photo emission spectra~\cite{ARPES_main,ARPES_cuprates} or scanning tunneling spectroscopy~\cite{STS_highTc_SC,STS_CaC6}, an analytic continuation is required, which is an ill-conditioned problem.
As a result, 
a variety of analytic continuation methods have been developed, such as Nevanlinna analytic continuations (NAC)~\cite{Neva1,Neva2}, the Padé approximation \cite{Pade_baker,lpade,Pade_beach}, Maximum Entropy (MaxEnt) formalism \cite{MaxEnt,MaxEnt_Bryan,MaxEnt_Kernel,Creffield95,Beach04,Gunnarsson10,Bergeron16,Levy17,Gaenko17,Rumetshofer19,Sim18}, stochastic analytic continuation \cite{Shao23,Sandvik98,Mishchenko00,Gunnarsson07,Fuchs10,Goulko17,Otsuki17,Krivenko19}, genetic algorithms, and machine learning \cite{Genetic,ML1,ML2} or causal projection methods \cite{PES}.

A highly topical and prominent use case for such a workflow is the state-of-the-art theoretical description of conventional superconductivity within Migdal-Eliashberg (ME) theory \cite{Migdal,Eliashberg,Marsiglio_ME}, as for instance implemented in the EPW package~\cite{epwcode2016,epwcode2023} of the Quantum ESPRESSO software~\cite{QE2009,QE2017}, which allows for the \textit{ab initio} computation of important properties of the superconducting phase, most notably the superconducting gap function $\Delta_{n\mathbf{k}}(i\omega_j)$, and thus \Tc{}, on the imaginary frequency axis. 

Many of the analytic continuation methods mentioned above require the \textit{causal} Green's function in imaginary time or frequency. This is a problem in ME formalism, where the electron-phonon mediated self-energy $\hat{\Sigma}_{n\mathbf{k}}$ and therefore the Green's function is split into three fundamental complex functions:
the already mentioned superconducting gap function $\Delta_{n\mathbf{k}}(i\omega_j)$, the mass renormalization function $Z_{n\mathbf{k}}(i\omega_j)$ and the energy shift function $\chi_{n\mathbf{k}}(i\omega_j)$. It is important to have access to those in real frequency space, which is why current implementations of analytic continuation in the EPW package are restricted to methods that are able to continue these three functions directly. This is the case for the computationally lightweight Padé approximation \cite{lpade} and a computationally very expensive iterative procedure \cite{lacon} specific to ME theory. The latter is generally not feasible in most scenarios due to numerical cost, and the former is known to often have numerical artifacts which break the non-negative causality condition of spectral functions.

In this work, we will demonstrate how to reconstruct the complex functions $\Delta_{n\mathbf{k}}(\omega)$, $Z_{n\mathbf{k}}(\omega)$ and $\chi_{n\mathbf{k}}(\omega)$ in real frequency space by analytically continuing the components of the Nambu-Gor'kov Green's function, and we further introduce an efficient general analytic continuation workflow for ME theory.
Importantly, this workflow can be used with any analytic continuation method that requires causal Green's functions, therefore giving us access to many more methods without any loss of information for the complex functions $\Delta_{n\mathbf{k}}(\omega)$, $Z_{n\mathbf{k}}(\omega)$ and $\chi_{n\mathbf{k}}(\omega)$. Due to symmetry, only two analytic continuations are required, which is one less than previously needed for full-bandwidth calculations with $\chi_{n\mathbf{k}}(\omega) \neq 0$~\cite{epwcode2023,lucrezi2024}. 

In this work, we employ the recently proposed NAC \cite{Neva1,Neva2,Neva_julia}, which is very well suited for continuing high-quality data and is able to reconstruct the spectral function with a high level of accuracy. We introduce a full implementation 
for both the isotropic and the anisotropic ME equations in the EPW package. We find that this method works exceptionally well for obtaining superconducting properties accurately, while keeping numerical costs comparable to the Padé approximation. NAC is numerically much more stable and, due to its analytic approach, always fulfills the causality conditions of normal spectral functions. 

\section{Theory}

\subsection{Nevanlinna Analytic Continuation (NAC)}

We start with a brief summary of the NAC as employed for our purposes. For a more detailed description of the method, we refer the reader to Refs.~\cite{Neva1,Neva2,Neva_julia}. 

Our goal is to obtain the experimentally measurable spectral function $A_{n\mathbf{k}}(\omega)$, which is proportional to the imaginary part of the retarded Green's function $G_{n\mathbf{k}}^\textrm{ret}$,
\begin{equation}
    A_{n\mathbf{k}}(\omega) = -\frac{1}{\pi}\,\textrm{Im}\,G_{n\mathbf{k}}^\textrm{ret}(\omega+i\eta),
    \label{eq:A_def}
\end{equation}
where $n$ and $\mathbf{k}$ are band and momentum indices, respectively. 
Due to the existence of poles on the real axis, the retarded Green's function is evaluated slightly above the real axis, described by the infinitesimal positive number $\eta$.
A causal spectral function such as the one in Eq.~(\ref{eq:A_def}) needs to fulfill two causality conditions,
\begin{equation}
    A_{n\mathbf{k}}(\omega) \geq 0 \qquad \textrm{and} \quad \int_{-\infty}^{\infty}\!\mathrm{d}\omega\, A_{n\mathbf{k}}(\omega) = 1.
    \label{eq:causal_cond}
\end{equation}
We want to note at this point that NAC can still be used in cases where normalization is not fulfilled, as will be the case for the auxiliary Green's function introduced in section~\ref{ch:NAC_with_ME}.

To obtain the retarded Green's function $G_{n\mathbf{k}}^\textrm{ret}(\omega+i\eta)$ from the imaginary frequency solution, analytic continuation in the form $G_{n\mathbf{k}}(i\omega_j) \rightarrow G_{n\mathbf{k}}^\textrm{ret}(\omega+i\eta)$ is required. This problem, however, is ill-conditioned, i.e. small errors in the Matsubara Green's function $G_{n\mathbf{k}}(i\omega_j)$ result in large errors in the retarded Green's function $G_{n\mathbf{k}}^\textrm{ret}(\omega+i\eta)$, regardless of the method one uses to analytically continue. The reason for that can be made apparent by considering
the kernel $K(\tau,\omega)$ of the spectral representation of the Green's function,
\begin{equation}
    G_{n\mathbf{k}}(\tau) = \int\!\mathrm{d}\omega \underbrace{\frac{e^{-\omega\tau}}{1+e^{-\omega\beta}}}_{K(\tau,\omega)} A_{n\mathbf{k}}(\omega),
    \label{eq:Kernel}
\end{equation}
which becomes exponentially small for high frequencies \cite{MaxEnt,MaxEnt_Kernel}. 
Determining the spectral function boils down to computing the inverse of the kernel, where for high frequencies divisions by very small numbers occur, resulting in small errors in the data having a huge impact on the result. Thus, the inverse of the kernel cannot be used to obtain the spectral function in practice.

\begin{figure}[t]
    \centering    
    \includegraphics[width=0.37\textwidth]{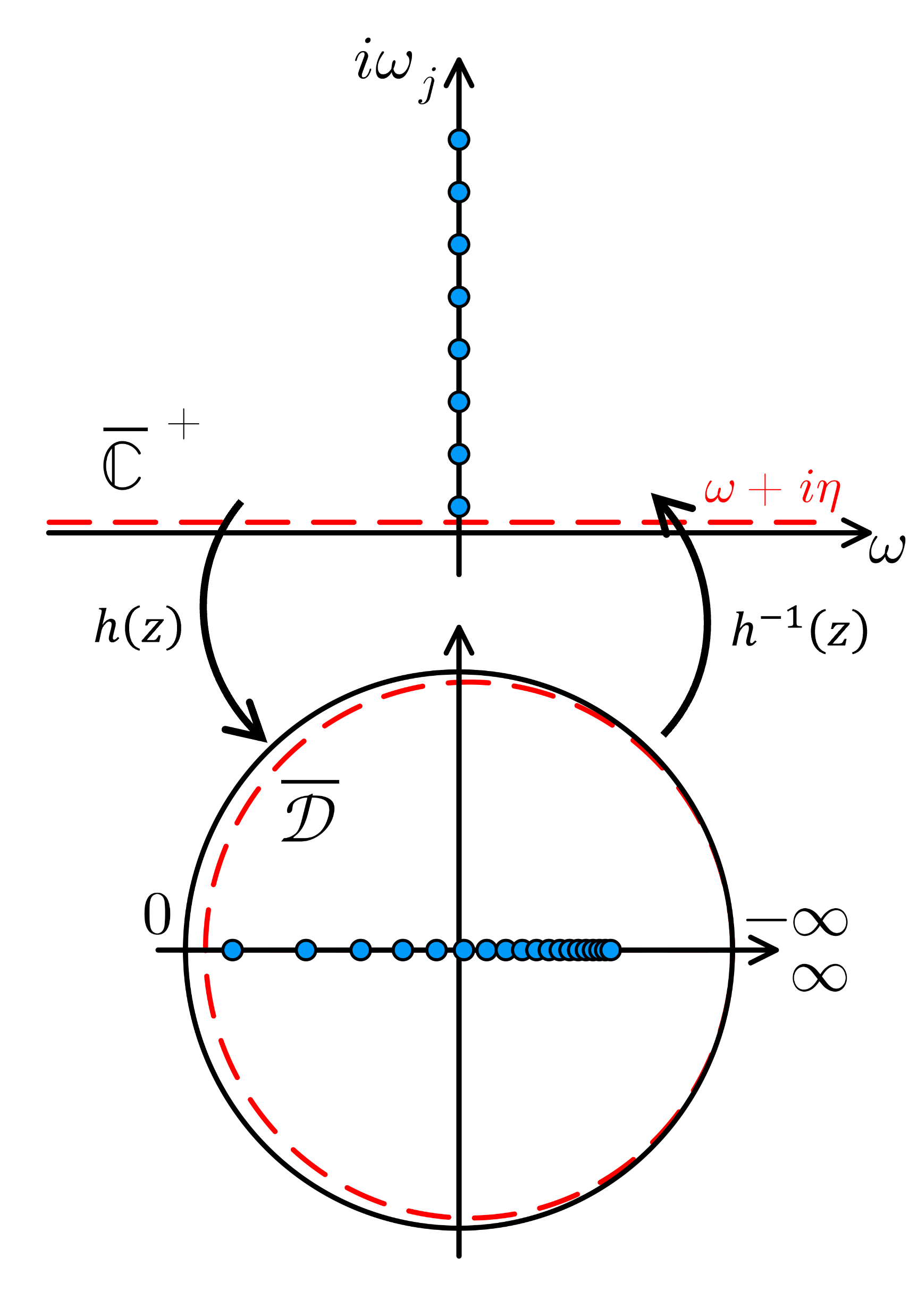}
    \caption{Sketch of the NAC procedure. Matsubara frequencies $i\omega_j$ (blue dots) and the Green's function $G_{n\mathbf{k}}(i\omega_j)$ are transformed from the positive complex plane $\overline{\mathbb{C}^+}$ to the unit circle $\overline{\mathcal{D}}$ using the M\"obius transformation $h(z)$. Once the Schur algorithm is performed, the function $\theta(z)$ is evaluated at $\omega+i\eta$ and transformed back to the complex plane using the inverse transformation $h^{-1}(z)$, to obtain the retarded Green's function.} 
    \label{fig:Moebius_trafo}
\end{figure}

The main interpolation procedure in NAC works according to the Schur algorithm \cite{Schur1,Schur2}, performed on the unit circle. In particular, to construct Nevanlinna interpolants, the first step is to M\"obius transform \cite{caratheodory} the Green's function to a contractive function on the unit circle with 
$\theta(i\omega_j) = (-G_{n\mathbf{k}}(i\omega_j) - i) / (-G_{n\mathbf{k}}(i\omega_j) + i)$,
as can be appreciated in Fig.~\ref{fig:Moebius_trafo}. We then use the following recursive formula from the Schur algorithm,
\begin{equation}
    \theta(z)[z;\theta_{M}(z)] = \frac{a(z)\theta_{M}(z)+b(z)}{c(z)\theta_{M}(z)+d(z)},
    \label{eq:Schur}
\end{equation}
with
\begin{equation}
    \begin{pmatrix}
        a(z) & b(z)\\
        c(z) & d(z)
    \end{pmatrix}
    = \prod_{j = 1}^{M}
    \begin{pmatrix}
        \frac {z-i\omega_j}{z+i\omega_j} & \theta_{j-1}(i\omega_j)\\
        \frac {z-i\omega_j}{z+i\omega_j} [\theta_{j-1}(i\omega_j)]^* & 1
    \end{pmatrix},
\end{equation}
to obtain analytic continuation on the unit circle $\theta(z)$. 

By construction, $\theta_0(z)=\theta(z)$ interpolates the Green's function for all Matsubara frequencies $j=1,2,...,M$, while each subsequent contractive function $\theta_1(z),...,\theta_M(z)$ interpolates one less point than the previous one. As a result, the final function $\theta_M$ is an unconstrained arbitrary contractive function on the unit circle. For discrete spectral functions of Lorentzian shape, the method works well regardless of the choice of $\theta_M(z)$, which is, for example, the case for most isotropic ME solutions. 

The simplest approach is using a constant $\theta_M(z)=0$, called the free solution.
However, in the case of smooth and rather featureless spectral functions, using a constant $\theta_M(z)$ can result in rapidly oscillating functions, 
in which case optimization of $\theta_M(z)$ can be required. We found that the free solution works very well in isotropic ME calculations, as discussed in section~\ref{ch:Results}. However, in the case of anisotropic ME theory, individual spectral functions will usually oscillate for $\theta_M(z)=0$. Optimization of $\theta_M(z)$ is possible~\cite{Neva1}, but will not be considered in this work and is subject of future studies.

Finally, to obtain the desired $G_{n\mathbf{k}}^\textrm{ret}(\omega+i\eta)$, the function $\theta(z)$ in Eq.~(\ref{eq:Schur}) is evaluated at $\omega+i\eta$ (red dashed line) and then transformed back to the complex plane using the inverse M\"obius transformation (see Fig.~\ref{fig:Moebius_trafo}).

One big advantage of NAC is that we can use a generalization of the Pick criterion \cite{Pick1,Pick2} to avoid the errors from higher frequency Matsubara points while also significantly improving numerical efficiency of the method. With the M\"obius transformation $h(z)=(z-i)/(z+i)$, the criterion is fulfilled if the matrix
\begin{equation}
    \left[ \frac{1-\theta(i\omega_j)\theta(i\omega_j)^*}{1-h(i\omega_i)h(i\omega_j^*)} \right]_{i,j} \quad i,j = 1,2,...,M
\end{equation}
is positive semi-definite, in our case verified by using Cholesky decomposition. When the Pick criterion is fulfilled, it is guaranteed that a solution to the Schur algorithm exists, which should always be the case for a noise-free imaginary solution. In practice, however, higher frequency points are much noisier, as has been discussed for the kernel in Eq.~(\ref{eq:Kernel}), so they will generally not fulfill the Pick criterion. As a result, discarding frequencies that don't fulfill the criterion reduces noise and improves numerical speed of the method.

\subsection{Using NAC with Migdal-Eliashberg Theory}
\label{ch:NAC_with_ME}

Contrary to a continued fraction approach like in the Padé approximation, NAC requires causal Green's functions to be employed. The Green's function we want to access in real frequency space using analytic continuation is the generalized Green's function $\hat{G}_{n\mathbf{k}}(\tau)$ from the Nambu-Gor'kov formalism~\cite{Nambu1,Nambu2},

\begin{equation}
    \hat{G}_{n\mathbf{k}}(\tau) = - 
    \begin{bmatrix}
        \langle T_\tau \hat{c}_{n\mathbf{k}\uparrow}(\tau) \hat{c}_{n\mathbf{k}\uparrow}^\dag(0) \rangle & 
        \langle T_\tau \hat{c}_{n\mathbf{k}\uparrow}(\tau) \hat{c}_{n-\mathbf{k}\downarrow}(0) \rangle \\ 
        \langle T_\tau \hat{c}_{n-\mathbf{k}\downarrow}^\dag(\tau) \hat{c}_{n\mathbf{k}\uparrow}^\dag(0) \rangle & 
        \langle T_\tau \hat{c}_{n-\mathbf{k}\downarrow}^\dag(\tau) \hat{c}_{n-\mathbf{k}\downarrow}(0) \rangle
    \end{bmatrix},
    \label{eq:Nambu_tau}
\end{equation}
where the main diagonal elements are the normal state Green's functions and the off-diagonal elements are the anomalous Green's functions, which describe Cooper pairs~\cite{Cooper1,Cooper2}.
Although the definition of the Nambu-Gor'kov matrix is different to a general matrix-valued Green's function, 
it is possible to transform one to the other using a spin-dependent particle-hole transformation such as $\hat{c}_{n-\mathbf{k}\downarrow} \rightarrow \hat{c}_{n-\mathbf{k}\downarrow}^\dag$. As a result, analytic continuation methods for matrix-valued Green's functions do work for the Nambu-Gor'kov matrix. However, due to symmetry properties of both the main and off-diagonal elements of $\hat{G}_{n\mathbf{k}}$, as well as its small size, an approach using auxiliary Green's functions as presented in this paper is much more efficient. As we will demonstrate later, only two analytic continuations are necessary to obtain all the components of the Nambu-Gor'kov matrix.

Due to the periodicity of the imaginary time Green's function, it can be expanded to imaginary Matsubara frequency space using a Fourier series of the form
\begin{align}
    \hat{G}_{n\mathbf{k}}(\tau) = \frac{1}{\beta}\sum_{n=-\infty}^{\infty}e^{-i\omega_j\tau} \hat{G}_{n\mathbf{k}}(i\omega_j),
\end{align}
and the Nambu-Gor'kov matrix in imaginary frequency space can be written as
\begin{equation}
    \hat{G}_{n\mathbf{k}}(i\omega_j) =  
    \begin{bmatrix}
        G^{11}_{n\mathbf{k}}(i\omega_j) & F_{n\mathbf{k}}(i\omega_j) \\ 
        F_{n\mathbf{k}}^*(i\omega_j) & G^{22}_{n\mathbf{k}}(i\omega_j)
    \end{bmatrix}.
    \label{eq:Nambu_iomega}
\end{equation}

One can explicitly show that this is fulfilled for the negative of the main diagonal elements of the Nambu-Gor'kov matrix using the Lehmann representation
\begin{equation}
    G^{11}_{n\mathbf{k}}(i\omega_j) = \frac{1}{Z} \sum_{mm'} \frac{\langle m'|\hat{c}_{n\mathbf{k}\uparrow}|m\rangle \langle m|\hat{c}_{n\mathbf{k}\uparrow}^\dag|m'\rangle}{i\omega_j + E_{m'} - E_m} (e^{-\beta E_m} + e^{-\beta E_{m'}})
\end{equation}
\noindent where $Z$ is the partition function, and $E_m$ and $E_{m'}$ are the eigenenergies of the states $|m\rangle$ and $|m'\rangle$. Identifying the positive term $K = Z^{-1} |\langle m|\hat{c}_{n\mathbf{k}\uparrow}^\dag|m'\rangle|^2 (e^{-\beta E_m} + e^{-\beta E_{m'}}) \ge 0$ and using $z=x+iy$, with $y>0$, we find
\begin{equation}
    \textrm{Im}[-G_{n\mathbf{k}}(i\omega_j)] = \sum_{mm'} \frac{Ky}{(x + E_{m'} - E_m)^2 + y^2} \ge 0,
\end{equation}
verifying the correct mapping $\mathcal{N}: \mathbb{C}^+ \rightarrow \overline{\mathbb{C}^+}$~\cite{Neva1,Neva2}. The prove for $G^{22}_{n\mathbf{k}}(i\omega_j)$ follows the same lines. 

The situation for the anomalous Green's function $F_{n\mathbf{k}}(i\omega_j)$ in the off-diagonal elements of Eq.~(\ref{eq:Nambu_iomega}) however is different, as the product of the matrix elements $\langle m'|\hat{c}_{n\mathbf{k}\uparrow}|m\rangle \langle m|\hat{c}_{n\mathbf{k}\uparrow}|m'\rangle$ cannot be rewritten as an absolute value. This means that the spectral function of the anomalous Green's function does not fulfill the causality condition of strictly non-negative spectral weight, 
and a direct analytic continuation using Nevanlinna functions is not possible.
This can also be seen from the anomalous spectral function $A^\textrm{an}_{n\mathbf{k}}(\omega)$ which is defined as usual from the imaginary part of $F_{n\mathbf{k}}(\omega)$. Due to the anti-commutation of the creation and annihilation operators with themselves, the normalization $\int_{-\infty}^{\infty}\!\mathrm{d}\omega\, A^\textrm{an}_{n\mathbf{k}}(\omega) = 0$ is fulfilled by definition.

Nevertheless, we can obtain the spectral functions of the anomalous Green's functions $F_{n\mathbf{k}}(i\omega_j)$ by constructing auxiliary Green's functions that can be continued with NAC. 
In accordance with Refs. \cite{auxGF_main,mixed_op_Gull} we define a mixed operator of the form
\begin{align}
    \hat{a}_{n\mathbf{k}}(\tau) = \hat{c}_{n\mathbf{k}\uparrow}(\tau) + \hat{c}_{n-\mathbf{k}\downarrow}^\dag(\tau).
\end{align}
The auxiliary Green's function $G^\textrm{aux}_{n\mathbf{k}}(\tau) = -\langle T_\tau \hat{a}_{n\mathbf{k}}(\tau) \hat{a}_{n\mathbf{k}}^\dag(0) \rangle$ is then guaranteed to have a spectral function with constant sign. Now, by rewriting the auxiliary Green's function in terms of the original creation and annihilation operators, one obtains
\begin{align}
    G^\textrm{aux}_{n\mathbf{k}}(\tau) = &- \langle T_\tau \hat{c}_{n\mathbf{k}\uparrow}(\tau) \hat{c}_{n\mathbf{k}\uparrow}^\dag(0) \rangle - \langle T_\tau \hat{c}_{n\mathbf{k}\uparrow}(\tau) \hat{c}_{n-\mathbf{k}\downarrow}(0) \rangle \\&- \langle T_\tau \hat{c}_{n-\mathbf{k}\downarrow}^\dag(\tau) \hat{c}_{n\mathbf{k}\uparrow}^\dag(0) \rangle - \langle T_\tau \hat{c}_{n-\mathbf{k}\downarrow}^\dag(\tau) \hat{c}_{n-\mathbf{k}\downarrow}(0) \rangle,
\end{align}
or, in other words, the sum over all the components of the Nambu-Gor'kov matrix. In Matsubara frequency space, the auxiliary Green's function is then given by
\begin{align}
    G^\textrm{aux}_{n\mathbf{k}}(i\omega_j) = \; &G^{11}_{n\mathbf{k}}(i\omega_j) + F_{n\mathbf{k}}(i\omega_j) \; + \nonumber\\ &F^*_{n\mathbf{k}}(i\omega_j) + G^{22}_{n\mathbf{k}}(i\omega_j).
    \label{eq:Gaux_iw_general}
\end{align}
In ME theory, the components of the Nambu-Gor'kov matrix from Eq. (\ref{eq:Nambu_iomega}) have the form
\begin{align}
    G^{11}_{n\mathbf{k}}(i\omega_j) &= -\frac{i\omega_j Z_{n\mathbf{k}}(i\omega_j) + \varepsilon_{n\mathbf{k}} + \chi_{n\mathbf{k}}(i\omega_j)}{\Theta_{n\mathbf{k}}(i\omega_j)} \\
    F_{n\mathbf{k}}(i\omega_j) &= F^*_{n\mathbf{k}}(i\omega_j)=-\frac{\Delta_{n\mathbf{k}}(i\omega_j) Z_{n\mathbf{k}}(i\omega_j)}{\Theta_{n\mathbf{k}}(i\omega_j)} \\
    G^{22}_{n\mathbf{k}}(i\omega_j) &= -\frac{i\omega_j Z_{n\mathbf{k}}(i\omega_j) - \varepsilon_{n\mathbf{k}} - \chi_{n\mathbf{k}}(i\omega_j)}{\Theta_{n\mathbf{k}}(i\omega_j)}
\end{align}
with
\begin{align}
    \Theta_{n\mathbf{k}}(i\omega_j) = &[\omega_j Z_{n\mathbf{k}}(i\omega_j)]^2 + [\varepsilon_{n\mathbf{k}} + \chi_{n\mathbf{k}}(i\omega_j)]^2 + \nonumber\\ &[\Delta_{n\mathbf{k}}(i\omega_j) Z_{n\mathbf{k}}(i\omega_j)]^2.
\end{align}
where we used the shorthand notation $\varepsilon_{n\mathbf{k}} = \epsilon_{n\mathbf{k}} - \epsilon_{F}$. Then the auxiliary Green's functions $G^\textrm{aux}_{n\mathbf{k}}(i\omega_j)$ in Matsubara frequency space is given by
\begin{align}
    G^\textrm{aux}_{n\mathbf{k}}(i\omega_j) = -\frac{2}{\Theta_{n\mathbf{k}}(i\omega_j)} \Big[ \Big. &i\omega_j Z_{n\mathbf{k}}(i\omega_j) + \Delta_{n\mathbf{k}}(i\omega_j) Z_{n\mathbf{k}}(i\omega_j) \Big. \Big].
    \label{eq:Gaux_iw_ME}
\end{align}

In contrast to the anomalous spectral function, the corresponding auxiliary spectral function $A^\textrm{aux}_{n\mathbf{k}}(\omega)$ is normalized to two, 
i.e. $\int_{-\infty}^{\infty}\!\mathrm{d}\omega\, A^\textrm{aux}_{n\mathbf{k}}(\omega) = 2$, and does fulfill the causality condition $A^\textrm{aux}_{n\mathbf{k}}(\omega) \geq 0$, allowing the application of NAC to $G^\textrm{aux}_{n\mathbf{k}}(i\omega_j)$.

The symmetry $G^{11}_{n\mathbf{k}}(-z)=-[G^{22}_{n\mathbf{k}}(z)]^*$
is fulfilled in both real and imaginary frequency space, which we can use to avoid explicit analytic continuation of $G^{22}_{n\mathbf{k}}$. As a result, we only need two analytic continuations to obtain all the components of the Nambu-Gor'kov matrix in real frequency space, one for $G^{11}_{n\mathbf{k}}$ and one for $G^\textrm{aux}_{n\mathbf{k}}$. 
The analytically continued anomalous Green's functions $F_{n\mathbf{k}}(\omega)$ is then given by 
\begin{align}
    F_{n\mathbf{k}}(\omega) = \frac{1}{2} \Big[ \Big. G^\textrm{aux}_{n\mathbf{k}}(\omega) - G^{11}_{n\mathbf{k}}(\omega) - G^{22}_{n\mathbf{k}}(\omega) \Big. \Big],
    \label{eq:F_from_aux}
\end{align}
where we do not explicitly write the small imaginary part $\eta$ to keep notations light.

We can now finally reconstruct all functions in ME formalism by using the Nambu-Gor'kov Green's function in real frequency space
\begin{align}
   \hat{G}_{n\mathbf{k}}(\omega) = \frac{\omega Z_{n\mathbf{k}}(\omega)\hat{\tau}_0 + [\varepsilon_{n\mathbf{k}} + \chi_{n\mathbf{k}}(\omega)]\hat{\tau}_3 + \Delta_{n\mathbf{k}}(\omega) Z_{n\mathbf{k}}(\omega)\hat{\tau}_1} {[\omega Z_{n\mathbf{k}}(\omega)]^2 - [\varepsilon_{n\mathbf{k}} + \chi_{n\mathbf{k}}(\omega)]^2 - [\Delta_{n\mathbf{k}}(\omega)Z_{n\mathbf{k}}(\omega)]^2}.
   \label{eq:Nambu_real}
\end{align}

Here, $\hat{\tau}_j$ are the Pauli matrices. By solving Eq. (\ref{eq:Nambu_real}) like a nonlinear system of equations, we find for the gap function

\begin{align}
    \Delta_{n\mathbf{k}}(\omega) = \frac{2\omega F_{n\mathbf{k}}(\omega)}{G^{11}_{n\mathbf{k}}(\omega) + G^{22}_{n\mathbf{k}}(\omega)},
    \label{eq:Delta_fromG}
\end{align}
in agreement with Refs.~\cite{Gap_from_Nambu,mixed_op_Gull, auxGF_main}. For the remaining complex functions in ME formalism, we have the following two equations
\begin{align}
    Z_{n\mathbf{k}}(\omega) = - \frac{G^{11}_{n\mathbf{k}}(\omega)+G^{22}_{n\mathbf{k}}(\omega)}{2\omega\left[[F_{n\mathbf{k}}(\omega)]^2-G^{11}_{n\mathbf{k}}(\omega)G^{22}_{n\mathbf{k}}(\omega)\right]},
    \label{eq:Z_fromG}
\end{align}
\begin{align}
    \chi_{n\mathbf{k}}(\omega) = \frac{-2\varepsilon_{n\mathbf{k}}[F_{n\mathbf{k}}(\omega)]^2+G^{11}_{n\mathbf{k}}(\omega)\left[2\varepsilon_{n\mathbf{k}}G^{22}_{n\mathbf{k}}(\omega)-1\right]+G^{22}_{n\mathbf{k}}(\omega)}{2\left[[F_{n\mathbf{k}}(\omega)]^2-G^{11}_{n\mathbf{k}}(\omega)G^{22}_{n\mathbf{k}}(\omega)\right]}.
    \label{eq:chi_fromG}
\end{align}
By obtaining the off-diagonal anomalous Green's function from the analytic continuation of the auxiliary Green's function in Eqs.~(\ref{eq:Gaux_iw_ME}) and (\ref{eq:F_from_aux}), there is no longer any loss in information, regardless of the analytic continuation method. This is the key idea of the generalized analytic continuation workflow we will introduce in more detail in section \ref{ch:Workflow}.

Eqs.~(\ref{eq:Delta_fromG}-\ref{eq:chi_fromG}) represent the set of self-consistent equations for the fully anisotropic ME theory. To speed up the calculations, two approximations are usually employed. (i) The first one is the isotropic limit, in which the gap function and the pairing are averaged over the Brillouin zone. This can be justified by considering that the superconducting gaps are found to be isotropic for most materials (MgB$_2$ being a notable exception) and that, in real materials, defects are always present, which tend to average out anisotropic gaps. (ii) Secondly, the density of states is considered constant as only electronic states within an energy window of the phonon energy around the Fermi level will experience renormalization due to electron-phonon effects. Considering only assumption (i), commonly referred to as the isotropic full-bandwidth ME equations, our framework simplifies to 
\begin{align}
    \Delta(\omega) = \frac{2\omega F(\omega)}{G^{11}(\omega)+G^{22}(\omega)},
\end{align}
\begin{align}
    Z(\omega) = - \frac{G^{11}(\omega)+G^{22}(\omega)}{2\omega\left[[F(\omega)]^2-G^{11}(\omega)G^{22}(\omega)\right]},
    \label{eq:Z_fromGiso}
\end{align}
\begin{align}
    \chi(\omega) = \frac{G^{22}(\omega)-G^{11}(\omega)}{2\left[[F(\omega)]^2-G^{11}(\omega)G^{22}(\omega)\right]}.
    \label{eq:chi_fromGiso}
\end{align}

Working with both assumptions, the energy shift function $\chi_{n\mathbf{k}}$ vanishes in imaginary and real frequency space. As a result the main diagonal components o the Nambu-Gor'kov matrix become identical, i.e. $G(\omega)=G_{n\mathbf{k}}^{11}(\omega)=G_{n\mathbf{k}}^{22}(\omega)$. So, the ME equations in our framework are further simplified to

\begin{align}
    \Delta(\omega) = \frac{\omega F(\omega)}{G(\omega)},
\end{align}
and
\begin{align}
    Z(\omega) = - \frac{G(\omega)}{\omega\left[[F(\omega)]^2-[G(\omega)]^2\right]}.
    \label{eq:Z_fromGiso_fsr}
\end{align}

One particularly interesting quantity of the superconducting state, apart from $\Delta$ and \Tc{} is the superconducting quasiparticle density of states, for the Fermi-surface restricted case given by
\begin{align}
    \frac{N_\textrm{S} (\omega)}{N_\textrm{F}} = -\frac{1}{\pi} \int_{-\infty}^{\infty}\!\mathrm{d}\epsilon_{n\mathbf{k}} \,\textrm{Im}\, G^{11}_{n\mathbf{k}}(\omega).
\end{align}

Although solving this integral numerically is possible from the $G_ {n\mathbf{k}}^{11}$ component, results will oscillate quite strongly beyond the initial gap for any method, as significantly finer $\mathbf{k}$-grids would be necessary to average over them properly. 

In the isotropic BCS-limit, where it is assumed that $\chi_{n\mathbf{k}}=0$ and $Z_{n\mathbf{k}}=1$ we have
\begin{align}
    \frac{N_\textrm{S} (\omega)}{N_\textrm{F}} = \sum_n \int\! \frac{\mathrm{d}\mathbf{k}}{\Omega_\textrm{BZ}} \frac{\delta(\epsilon_{n\mathbf{k}} - \epsilon_{F})}{N_\textrm{F}} \textrm{Re} \left[ \frac{\omega}{\sqrt{\omega^2 - \Delta^2_{n\mathbf{k}}(\omega)}} \right].
\end{align}
In other words, by using Eq. (\ref{eq:Delta_fromG}), we also have access to the quasiparticle density of states in the BSC-limit, which generally results in more pronounced peaks and smoother curves.

\subsection{Analytic Continuation Workflow for Migdal-Eliashberg Theory using Green's functions}
\label{ch:Workflow}
\begin{figure}[t]
    \centering
    \includegraphics[width=0.42\textwidth]{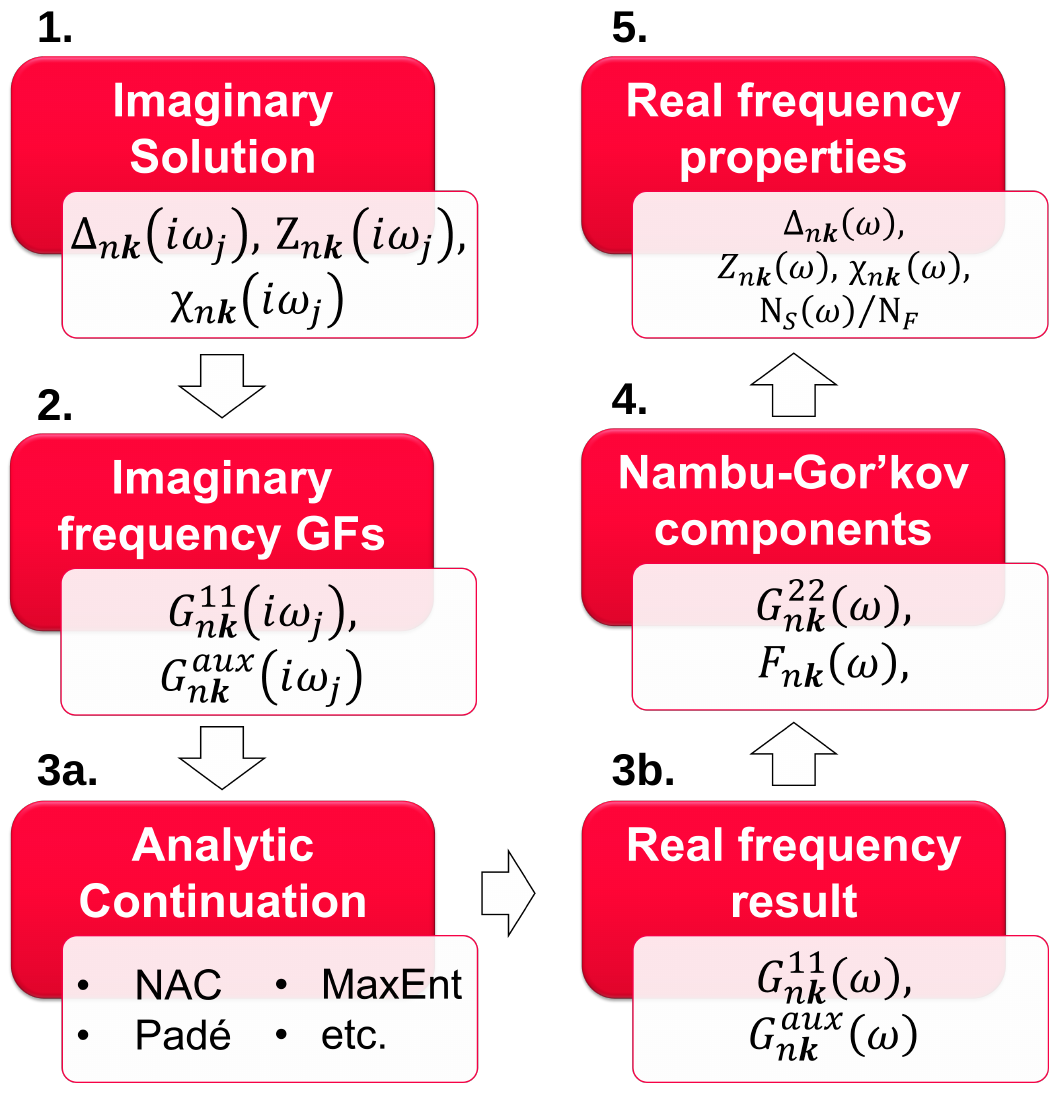}
    \caption{Flowchart for the analytic continuation procedure in ME theory using Green's functions.}
    \label{fig:Flowchart}
\end{figure}

To perform the analytic continuation of the complex functions $\Delta_{n\mathbf{k}}$, $Z_{n\mathbf{k}}$ and $\chi_{n\mathbf{k}}$ from ME formalism using causal Green's functions, we introduce the following generalized workflow as shown in Fig.~\ref{fig:Flowchart}. The steps are as follows:

\begin{enumerate}
    \item Start with the imaginary frequency solution from ME theory $\Delta_{n\mathbf{k}}(i\omega_j)$, $Z_{n\mathbf{k}}(i\omega_j)$ and $\chi_{n\mathbf{k}}(i\omega_j)$.
    \item Calculate the normal Green's function $G^{11}_{n\mathbf{k}}(i\omega_j)$, as well as the auxiliary Green's function $G^\textrm{aux}_{n\mathbf{k}}(i\omega_j)$ from Eq.~(\ref{eq:Gaux_iw_ME}).
    \item Perform the analytic continuation of these two Green's functions (a) to obtain them in real frequency (b).
    \item Determine the remaining components of the Nambu-Gor'kov matrix in real frequency, by using the symmetry $G^{22}_{n\mathbf{k}}(\omega)=-[G^{11}_{n\mathbf{k}}(-\omega)]^*$, and Eq. (\ref{eq:F_from_aux}) for the anomalous Green's function $F_{n\mathbf{k}}(\omega)$.
    \item Calculate any other real frequency properties from the obtained Nambu-Gor'kov matrix, as for instance the gap function $\Delta_{n\mathbf{k}}(\omega)$ and the quasiparticle density of states in the BCS-limit $\frac{N_\textrm{S} (\omega)}{N_\textrm{F}}$.
\end{enumerate}
This workflow can be used with any analytic continuation method. Even in the case of the Padé approximation, which initially does not require causal Green's functions, it is possible to restrict the initial continuous fraction assumption to the analytic continuation of causal Green's functions~\cite{Pade_cont_frac}
\begin{align}
    G_{n\mathbf{k}}(z) = \cfrac{a_1}{z+b_1-\cfrac{a_2}{z+b_2-\cfrac{a_2}{z+b_3-...}}},
    \label{eq:cont_frac}
\end{align}
where $a_j > 0$ and $b_j\in\,\mathbb{R}$. 

Using this causal Green's function approach for the Padé approximation could make the method numerically more stable within ME theory, but is not the focus of this work.

\section{Results}
\label{ch:Results}

\begin{figure}[t]
    \centering
    \includegraphics[width=0.45\textwidth]{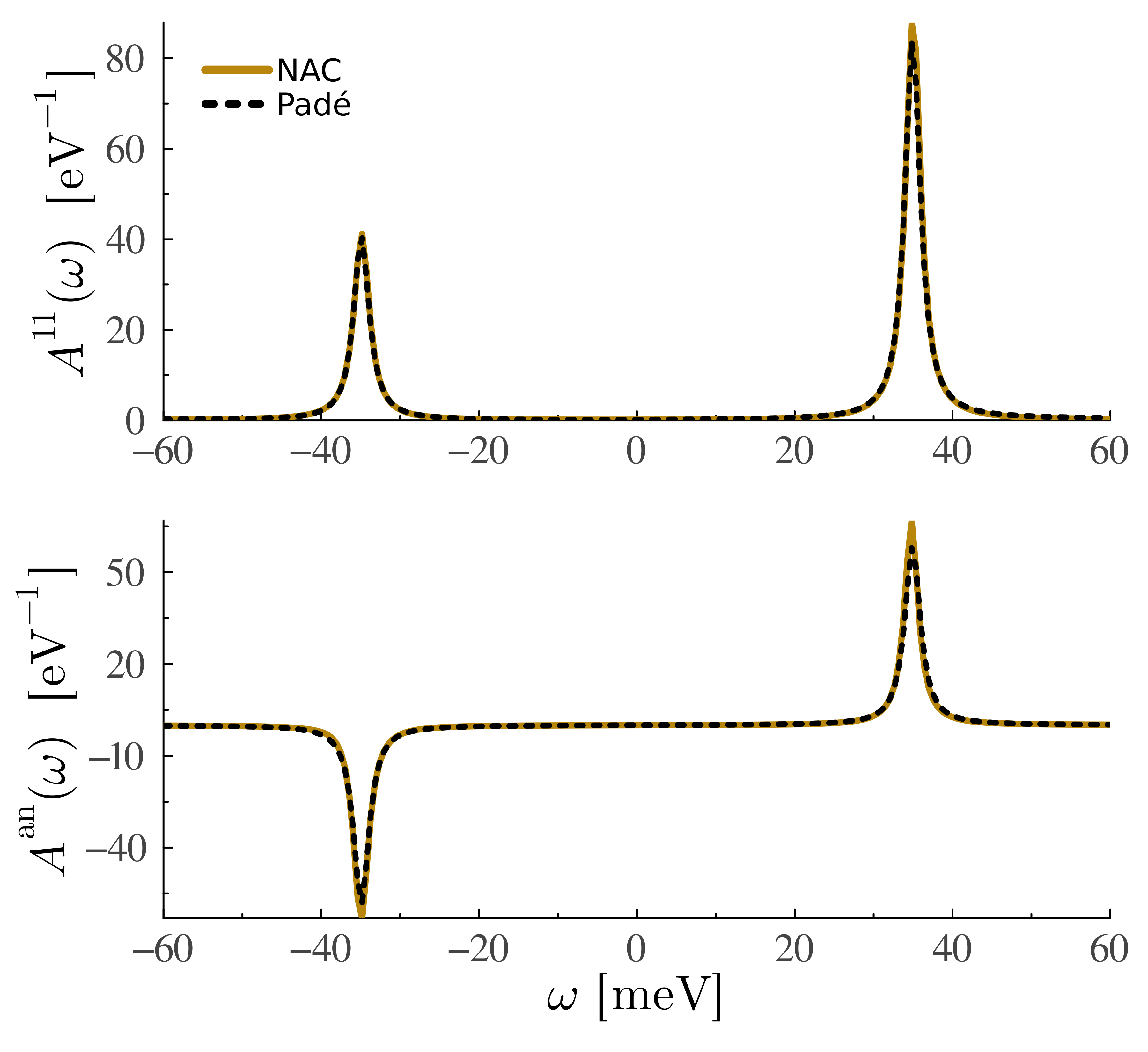}
    \caption{Normal and anomalous spectral functions for LaBeH$_8$ at a temperature of $T = 70$ K within the isotropic full-bandwidth approximation, obtained with EPW. This is the most common case, where both methods return the same result.
    }
    \label{fig:Fall1}
\end{figure}

We start by showing some results where NAC and the Padé approximation are in excellent agreement. First, for an isotropic full-bandwidth calculation of LaBeH$_8$, with a fine $\mathbf{k}$-grid of $30 \times 30 \times 30$ at a temperature of $T=70$\,K. We used default parameters for the Padé approximation as implemented in EPW. For NAC, we used the Pick criterion to determine the optimal number of Matsubara frequencies and a broadening parameter of $\eta=1$\,meV. The same broadening was added to the spectral function from the Padé approximation in post-processing to obtain Fig.~\ref{fig:Fall1}, where we show a comparison between the two methods for both the normal and anomalous spectral functions $A^{11}_{n\mathbf{k}}(\omega)$ and $A^{\textrm{an}}_{n\mathbf{k}}(\omega)$, respectively. The pole position, and therefore the quasiparticle energy renormalized by superconducting pairing, is located around $\omega=\pm 35$\,meV. NAC is done without optimization, i.e. by using a constant free parameter of $\theta_M(z)=0$. We found that this works very well for the isotropic approximation to ME theory, although small unphysical satellite peaks with a maximum of $1$\,eV$^{-1}$ to $2$\,eV$^{-1}$ can appear at higher frequencies. 
The most common outcome is this kind of agreement between NAC and the Padé approximation, which we found to be the case in about 95\% of our testing.

NAC can also be used for anisotropic calculations, however spectral functions of individual $n$ and $\mathbf{k}$ index often have oscillations for $\theta_M(z)=0$, meaning an optimization of the $\theta_M(z)$ parameter would be required. Work on this is ongoing and will be implemented in a future version. 

\begin{figure}[t]
    \centering
    \includegraphics[width=0.47\textwidth]{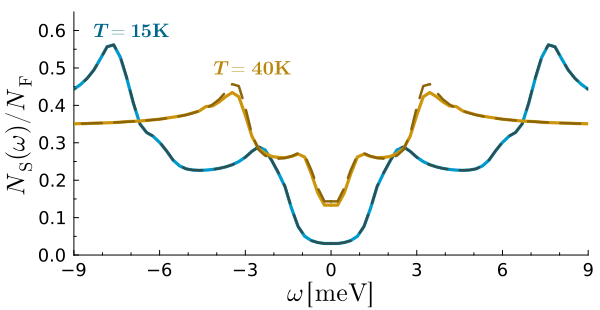}
    \caption{Quasiparticle density of states of MgB$_2$ for different temperatures within the anisotropic Fermi-surface restricted approximation, obtained with EPW. Comparison between Padé approximation (dashed lines) and NAC (solid lines) for two different temperatures.}
    \label{fig:qdosneva_MgB2}
\end{figure}

Nevertheless, we can use NAC to obtain excellent agreement in the quasiparticle density of states, demonstrated in Fig. \ref{fig:qdosneva_MgB2} for the well-studied two-gap superconductor MgB$_2$~\cite{Margine_ME,MgB2_origin_2gap}. Although slight oscillations persist for individual $n$ and $\mathbf{k}$ contributions when NAC is combined with anisotropic ME calculations, they effectively cancel out entirely in the quasiparticle density of states if the fine $\mathbf{k}$-grid is sufficiently dense. As a result we have almost perfect agreement between the two methods.
From Fig. \ref{fig:qdosneva_MgB2}, we can, therefore, identify the gaps that originate from the different electronic $\pi$ and $\sigma$ orbitals. For example, in our calculation for $T=15$\,K, we find a first gap of about $2\Delta_\pi=5$\,meV and a second gap of about $2\Delta_\sigma=16$\,meV. Finally, we obtained a critical temperature of \Tc{}$ = 42$\,K for our anisotropic calculation of MgB$_2$.

\begin{figure}[t]
    \centering
    \includegraphics[width=0.45\textwidth]{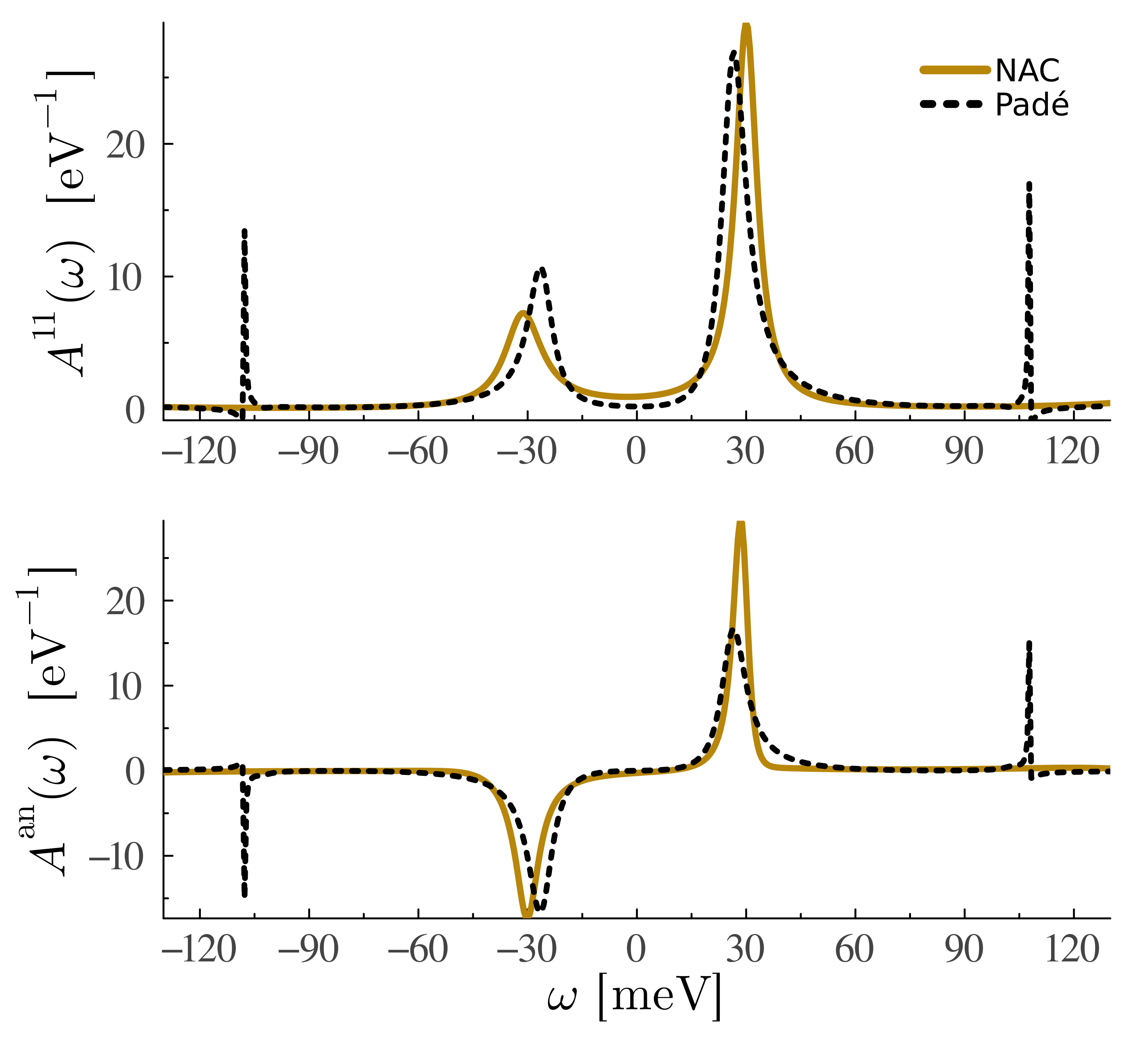}
    \caption{Normal and anomalous spectral functions for LaBeH$_8$ at a temperature of $T=130$ K within the isotropic full-bandwidth approximation obtained with EPW. Here, NAC is able to avoid the numerical error of the Padé approximation at around $\omega=\pm 110$\,meV.
    }
    \label{fig:Fall2}
\end{figure}

Beyond the most common 
cases with excellent agreement between NAC and the Padé approximation, we found two significant advantages of NAC: The first rather obvious one is the avoidance of non-causal negative $A^{11}_{n\mathbf{k}}(\omega)$ spectral functions as shown in Fig.~\ref{fig:Fall2}, demonstrating a situation where the Padé approximation exhibits non-physical numerical artifacts at about $\omega=\pm 110$\,meV. 
By definition, the spectral functions from NAC is always positive, making the method preferable for analytic continuations of simple Lorentzian peak shaped spectral functions, which is almost always the case for isotropic ME calculations.

\begin{figure}[t]
    \centering
    \includegraphics[width=0.45\textwidth]{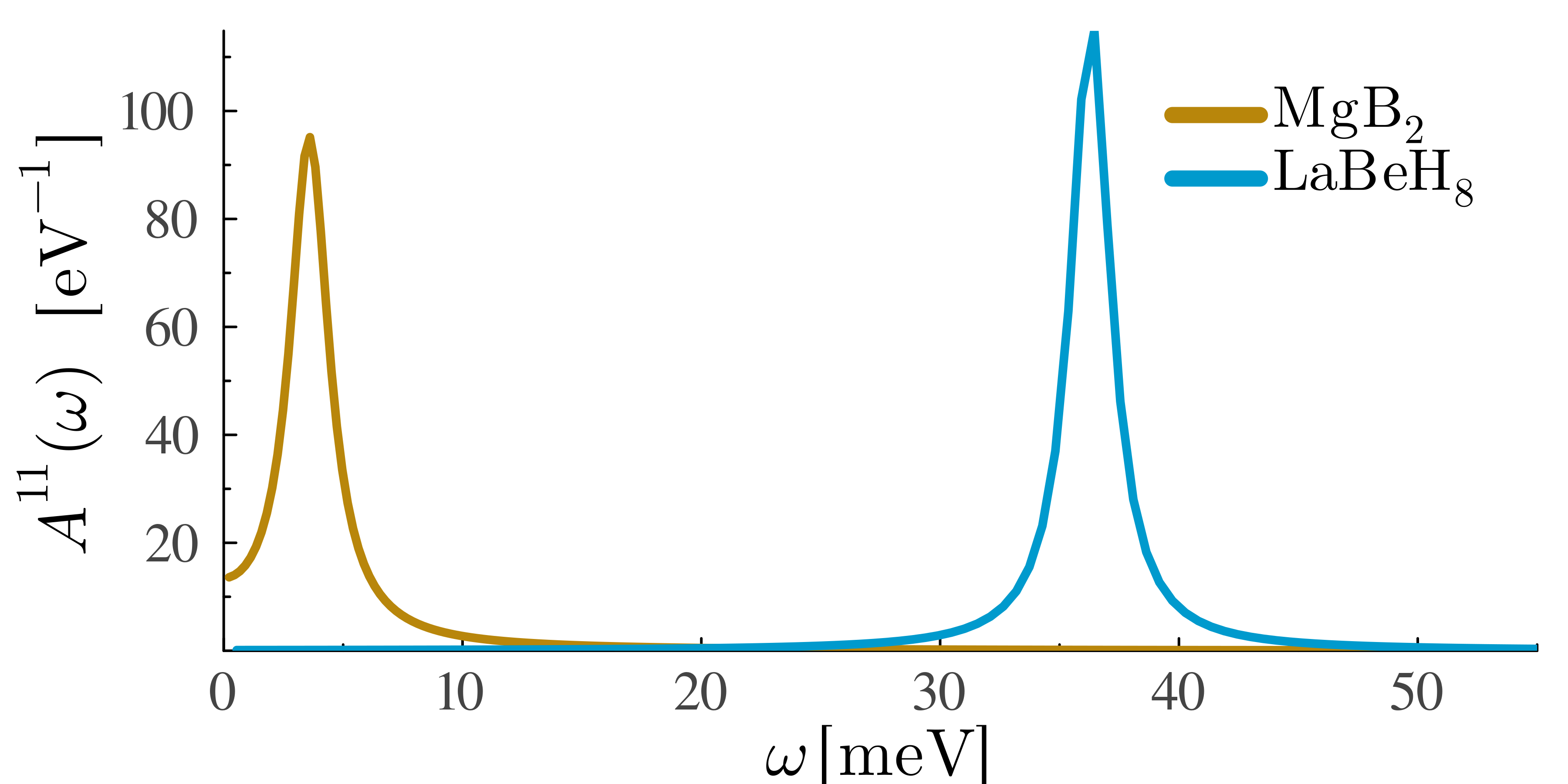}
    \caption{Normal spectral functions for MgB$_2$ and LaBeH$_8$ at a very low temperature of $T=0.05$\,K within the isotropic Fermi-surface restricted approximation, obtained with EPW. The Padé approximation failed in these cases.
    }
    \label{fig:Fall3}
\end{figure}

The second advantage is that NAC is able to analytically continue even in difficult cases where the Padé approximation fails entirely. Most common for isotropic calculations at very low temperatures, the computation of the Padé interpolants will run into \textit{division by zero} issues. 
The reason for this is that at low temperatures, the ME calculation requires a lot of Matsubara points, which then enter the Padé approximation and significantly increase the probability of failure. 
This can be appreciated in Fig.~\ref{fig:Fall3} for an isotropic calculation of MgB$_2$ and LaBeH$_8$ at a low temperature of $T=0.05$\,K. NAC is numerically much more stable by introducing quad precision arrays in the Schur algorithms of the method. We tested a similar quad precision approach for the Padé approximation; however, the numerical cost becomes much larger by approximately a factor of ten times more than NAC. The reason for this is the efficiency of the Pick criterion, which allows NAC to require much less Matsubara frequencies. However, even when using the same number of Matsubara frequencies for both methods, i.e. by not using the Pick criterion, NAC has never run into \textit{division by zero} issues for our testing, as numerical errors only appear in the form of the above-mentioned satellite peaks.

The numerical costs of NAC using quadruple precision are very similar to those of the Padé approximation using double precision. We reached this conclusion by comparing the runtimes of anisotropic calculations, as isotropic calculations are too brief to provide meaningful comparisons. However, it's important to note that the runtimes of anisotropic calculations depend highly on the chosen $\mathbf{k}$-grid, so the following estimates are intended to give a general sense of the performance. In our calculations for MgB$_2$, we observed typical CPU runtimes of approximately $50$s per temperature for NAC with quadruple precision, comparable to the runtimes for Padé approximation calculations using double precision. In contrast, using quadruple precision for the Padé approximation significantly increases the computation time to around $500$s.

\section{Conclusion}
In this work, we demonstrate a method to obtain
real-frequency properties from the analytically continued Nambu-Gor'kov matrix within the framework of Migdal-Eliashberg (ME) theory for superconductivity.
By leveraging the symmetry properties of the Green's function, our method requires only two analytic continuations to obtain all components in real frequency space: one for the normal and one for the auxiliary Green's function. This significantly broadens the scope for applying a variety of analytic continuation techniques that were previously inaccessible in the Migdal-Eliashberg framework, retaining full information of the real frequency complex functions explicitly, such as $\Delta_{n\mathbf{k}}(\omega)$, $Z_{n\mathbf{k}}(\omega)$, and $\chi_{n\mathbf{k}}(\omega)$. Our generalized workflow opens up the possibility for employing advanced methods like MaxEnt, stochastic analytic continuation, genetic algorithms, and causal projection in the modeling of superconducting materials, thereby enhancing the applicability and accuracy of the Migdal-Eliashberg formalism.

In particular, as a very powerful method for analytic continuation, we implemented the NAC into this framework.

Compared to the standard Padé approximation, our method offers two key advantages: (i) it effectively avoids non-physical negative values in the spectral function, and (ii) it provides improved numerical stability, enabling accurate analytic continuations at significantly lower temperatures than previously achievable. Despite these enhancements, the NAC method maintains comparable CPU runtimes, thanks to the efficiency of the Pick criterion and the streamlined analytic continuation workflow we introduced.

Using NAC with an unoptimized $\theta_M(z)$ parameter performs well for isotropic Migdal-Eliashberg calculations at lower frequencies. However, minor satellite peaks can emerge at higher frequencies due to random oscillations, indicating the need for careful optimization of the $\theta_M(z)$ parameter. Likewise, optimization can be necessary to mitigate oscillatory behavior and ensure accurate results when dealing with spectral functions from anisotropic calculations.

\section{Data availability}

The authors confirm that the data supporting the findings of this study are available within this article. The updated and newly developed code routines can be obtained upon request and will also be made available within a future EPW release.

\acknowledgements
We thank E. Gull, E. R. Margine, and H. Mori for their insightful discussions. Calculations have been performed on the dCluster and the lCluster of the Graz University of Technology and on the Vienna Scientific Cluster (VSC) under Project 71754. PNF acknowledges the S\~{a}o Paulo Research Foundation (FAPESP) under Grants 2020/08258-0 and 2021/13441-1.

\appendix

\section{Computational Details}
DFT and DFPT calculations were performed using the Quantum ESPRESSO (QE) software, employing norm-conserving Vanderbilt (ONCV) pseudopotentials \cite{ONCV} with a PBE-GGA exchange-correlation potential.

For $Fm\overline{3}m$-LaBeH$_8$ at $100$ GPa, we used a $12\times12\times12$ Monkhorst-Pack $\mathbf{k}$-grid, a $6\times6\times6$ Monkhorst-Pack $\mathbf{q}$-grid, a kinetic energy cutoff of $80$\,Ry, a smearing of $0.01$\,Ry, an electronic convergence threshold of $10^{-10}$\,Ry, and a phonon self-consistency threshold of $10^{-14}$\,Ry. 

After that, we used the Wannier implementation of EPW to interpolate electron-phonon matrix elements onto dense $\mathbf{k}$- and $\mathbf{q}$-grids. We used the "selected columns of the density matrix" (SCDM) method \cite{SCDM,SCDM_applic} to automatically wannierize our electronic band structure, using a frozen window of about $[-14$\,eV, $+5$\,eV] around the Fermi energy $E_F$. For the SCDM entanglement, we use an error function with the two parameters $\sigma=5$ and $\mu_c$ at the Fermi energy $E_F$. Then for the solution of the isotropic full-bandwidth ME equations, we used a coarse $\mathbf{k}$- and $\mathbf{q}$-grids of $6\times6\times6$ interpolated onto $30\times30\times30$ fine grids. We set the lower boundary of phonon frequency to $10$\,cm$^{-1}$, the width of the Fermi-surface window to 1\,eV, and the imaginary convergence threshold to $10^{-5}$\,Ry, all for a Morel-Anderson pseudopotential of $\mu^*=0.16$. The ME equations were solved while updating the chemical potential to keep the particle number constant, which requires a comparably high frequency cut-off of $10$\,eV for convergence. We used sparse sampling and analytically continued with the Padé approximation and NAC simultaneously, starting at a temperature of $60$\,K increased by steps of $10$\,K.

For MgB$_2$, we used 80\,Ry for the kinetic energy cutoff, 24$^3$ Monkhorst-Pack $\mathbf{k}$-grid, and a Methfessel-Paxton first-order spreading~\cite{MP-smearing} of 0.02\,Ry for Brillouin zone integration. The dynamical matrices and the variations of the self-consistent potentials were obtained on a 6$^3$ Monkhorst-Pack $\mathbf{q}$-grid with a phonon self-consistency threshold of $10^{-16}$\,Ry.

The electronic wave functions for the Wannier interpolation were calculated on a $\Gamma$-centered uniform $\mathbf{k}$-grid of 6$^3$ points. We used 5 Wannier functions for describing the band structure of MgB$_2$ around the Fermi level: two B-$p_z$-like states and 3 $sp^2$ hybridized orbitals associated with the Mg atom.

To solve the Eliashberg equations for MgB$_2$, we adopted the Fermi-surface restricted approximation~\cite{lucrezi2024} to keep our calculations consistent with the approach adopted in Ref.~\cite{Margine_ME}. We interpolated the electron-phonon matrix elements onto fine $\Gamma$-centered uniform grids containing 60$^3$-$\mathbf{k}$ and 30$^3$-$\mathbf{q}$ points. The Matsubara frequency cutoff is set to 1\,eV, approximately 10 times the maximum phonon frequency. The smearing parameters in the Dirac $\delta$ functions for electrons and phonons are set to $100$\,meV and $0.5$\,meV, respectively. The Pick criterion was used to determine the optimal number of Matsubara points, and the real frequency broadening of NAC was set to zero.

For the Coulomb pseudopotential $\mu^{*}$ of MgB$_2$, we utilized the $\mu^{*}$ calculated in Ref.~\cite{ferreira2024} according to the Morel-Anderson approximation at the $GW$ level. To rescale the value of $\mu^{*}$ used for solving the Allen-Dynes equation in Ref.~\cite{ferreira2024} to be used in the Eliashberg equation in the present work, we used the simple rule~\cite{pellegrini2024}
\begin{align}
    \dfrac{1}{\mu^{*}_{\text{Eliashberg}}} = \dfrac{1}{\mu^{*}_{\text{AD}}} + \ln{\left(\dfrac{\omega_{\text{ph}}}{\omega_{\text{c}}}\right)},
\end{align}
where $\omega_{\text{ph}}$ is the maximum of the phonon spectrum. By doing this, we obtained a $\mu^{*}$ of 0.11.

\section{EPW Input Flags}
We have implemented three new input flags that are all written into the input file of the EPW calculation. Although not necessary, it is good practice to perform analytic continuations using a restart calculation. We require the imaginary ME solution, so "limag=.true." has to be set. Then, to enable the NAC calculation, use the logical flag "lneva=.true." (default .false.), which can also be done in the same run as any other analytic continuation method. By default, the Pick criterion will be used to determine the optimal number of Matsubara points. However, if that is not desirable one can disable the criterion with "lpick=.false." (default .true.), where all Matsubara frequencies are used instead. Finally, we have implemented a flag for the real frequency broadening $\eta$ (in eV) of NAC, called "neva\_broad". By default, a small broadening of "neva\_broad=0.001", or $1$\,meV, is used.

\bibliographystyle{apsrev4-2}
\bibliography{refs}

\end{document}